\documentclass[traditabstract,twocolumn]{aa}

\usepackage{graphicx,color} 
\usepackage{txfonts}
\usepackage[]{natbib}

\begin{document}
\title{Discovery of an X-ray cavity near the radio lobes of Cygnus A
indicating previous AGN activity}
\author{Gayoung Chon\inst{1}, Hans B\"ohringer\inst{1}, 
Martin Krause\inst{2,1}, Joachim Tr\"umper\inst{1}}
\offprints{Gayoung Chon, gchon@mpe.mpg.de} 
\institute{$^1$ Max-Planck-Institut f\"ur extraterrestrische Physik, 
85748 Garching, Germany\\
$^2$ Excellence Cluster Universe, Technische Universit\"at 
M\"unchen, Boltzmannstrasse 2, 85748 Garching, Germany
}
\date{Submitted 3 May 2012;} 

\abstract
{Cygnus A harbours the nearest powerful radio jet of an 
Fanaroff-Riley (FR) class II radio 
galaxy in a galaxy cluster where the interaction of the jet with 
the intracluster medium (ICM) can be studied in detail. We use a 
large set of \emph{Chandra} archival data, \emph{VLA} and new 
\emph{LOFAR} observations to shed new light on the interaction 
of the jets with the ICM. We identify an X-ray cavity in the 
distribution of the X-ray emitting plasma in the region south of 
the Cyg A nucleus which has lower pressure than the surrounding 
medium. The \emph{LOFAR} and \emph{VLA} radio 
observations show that the cavity is filled with synchrotron 
emitting plasma. The spectral age and the buoyancy time of 
the cavity indicates an age at least as large as the current
Cyg A jets and not much larger than twice 
this time. We suggest that this cavity was created in a previous 
active phase of Cyg A when the energy output of the Active
Galactic Nucleus (AGN) was about two orders of magnitude less than today.
\keywords{X-rays: galaxies: clusters - Galaxies: clusters: intracluster medium - Radio continuum: galaxies - Galaxies: jets}
}
\authorrunning{Chon et al.} 
\titlerunning{X-ray cavity in the radio lobes of Cygnus A}
\maketitle
%

\section{Introduction}

AGN feedback in galaxy clusters has gained a lot of attention
in recent studies, e.g.~\cite{mcnamara}.
Most of the AGN in clusters are FR I radio sources. In the local Universe 
Cygnus A is an exception featuring a powerful FR II radio jet in a galaxy 
cluster where very energetic interaction effects of the jets with the 
intracluster medium (ICM) can be studied in detail, 
e.g.~\citet{carilli1996,clarke,kaiser,krause}. 
In this object we can observe jets of relativistic synchrotron 
emitting plasma plowing into the intracluster medium out 
to a distance of $\sim$100~kpc from the nucleus where they end 
in hotspots clearly seen at radio wavelengths, (e.g.~\citealt{alexander}) 
and X-ray images (e.g.~\citealt{harris}). 
The internally supersonic jet plasma is thermalised at these hotspots, 
which advance only at about 2000~km/s due to the extreme density 
contrast (\citealt{alexander_book}). 
The jet plasma flows away from the overpressured hotspots establishing
a backflow. Together with the adjacent ICM, 
this region defines an overpressured bubble which drives a shock into 
the undisturbed ICM, first seen by~\cite{carilli1988} in the centre of 
the Cyg A cluster.
%

Many aspects of Cyg A have been studied in both radio and X-rays,
e.g.~\citet{carilli1994,wilson2000,wilson2006,smith,yaji}.
Some progress has been made recently in modeling these jets with 
hydrodynamical simulations (\citealt{krause03,krause}, 
see also~\citealt{alexander_book}). 
The observed width of 
the radio lobes are only reproduced by very light 
jets, with jet densities of order $10^{-4}$ times the already tenuous 
surrounding X-ray plasma. This implies a low momentum flux and 
low hotspot advance speeds, comparable to the general 
advance speed of the bow shock driven into the surrounding ICM. Consequently, 
the bow shock assumes an almost spherical shape around the radio source 
with a wide region of shocked ICM adjacent to the radio lobes. 

In this paper we use most of the \emph{Chandra} archival data 
on Cyg A to study in detail the interaction effects in the region
near the AGN inside the radio lobes. In this region we discovered
an X-ray cavity very similar to those found in FR I radio sources
in clusters and investigated it in detail with the X-ray and radio data 
to unveil its physical properties and its origin, and 
the interpretation with a hydrodynamical simulation.

For distance related quantities we adopt a flat $\Lambda$-cosmology with 
$H_0$=70 km/s/Mpc, $\Omega_M$=0.3. For the redshift of Cyg A
of 0.0561, 1\arcsec corresponds to 1.09 kpc.

\section{\emph{Chandra} observations}

\begin{figure}
\begin{center}
\resizebox{\hsize}{!}{\includegraphics[width=6.5cm]{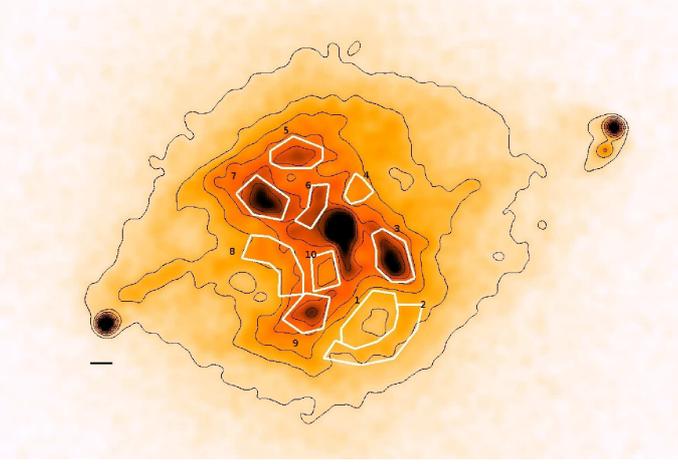}}
\end{center}
\caption{\emph{Chandra} image of Cygnus A with contours and 
ten study regions in white polygons with their ID numbers. 
The dimension of the image is 2.6\arcmin$\times$1.75\arcmin, 
and a black line of length 5\arcsec is marked below the eastern 
hotspot. The X-ray cavity, 1, is the polygon defining a region 
of low X-ray surface brightness. 
}
\label{fig:cimage}
\end{figure}

We used all available archival \emph{Chandra} ACIS-I observations for
Cyg A, which sum up to 200 ks of exposure. Standard data analysis 
was performed with CIAO 4.4\footnote{http://cxc.harvard.edu/ciao4.4/index.html} 
with calibration database (CALDB) 4.4.8. This includes flare cleaning 
by filtering light curves of source-free regions, VFAINT mode,
gain, and charge transfer inefficiency corrections. The final 
exposure after cleaning is slightly less than 190ks. 
We apply the same procedure to an ACIS blank sky observation, 
which was used as our background estimate after adjusting 
the normalisation. In our case different normalisations of the 
background do not significantly influence the results of
our analysis since the region of our interest has high source counts.
We used the 9-12.5 keV countrate to get the appropriate normalisation 
for the background exposure.
We used \emph{fluximage} to obtain flux and exposure maps after
background subtraction, and the combined map is shown in 
Fig.~\ref{fig:cimage} and~\ref{fig:radio} in the 0.5--2 keV band. 
Fig.~\ref{fig:cimage} shows a detailed image of Cyg A with
X-ray contours in an area of 2.6\arcmin$\times$1.75\arcmin.
There are ten regions defined by white polygons with their ID numbers, 
for which we determine the X-ray properties in the next section. 

\section{Properties of the X-ray Cavity}

Fig.~\ref{fig:cimage} shows the detailed X-ray surface brightness 
distribution in the shocked region of Cyg A. The shocked region
is contained in an ellipse surrounding two hotspots and the jets, 
and is extending outside the outermost contour in Fig.~\ref{fig:cimage}.
From radio observations we know that the energetic jet created
backflows, which are seen as diffuse lower surface brightness areas
starting near the hotspots towards the centre.
The central region away from the heads of the jets 
and backflows is expected to be in pressure equilibrium. 
An interesting region drawing an attention is the low surface brightness
feature numbered 1. This is the ``cavity'' we study in this paper
with a lower than expected thermal plasma pressure.

The goal of the X-ray analysis is then to extract spectra for the 
determination of temperatures, to calculate electron densities, $n_e$, 
and pressures to characterise the cavity and its vicinity to diagnose 
the pressure distribution. $n_e$ is derived from the X-ray luminosity, 
$L_x=\int dV n_e^2 \Lambda (T)$, and temperatures are determined
by fitting the observed spectrum in XSPEC with an absorbed MEKAL model.

We defined ten regions of roughly homogeneous surface brightness and
sufficient source counts in the centre of Cyg A around region 1.
For comparison four regions (3,5,7,9) were similarly defined as 
in~\citet{smith,wilson2006}. Our fitted temperatures from these four 
regions agree well with theirs within our uncertainties. 
We find the temperature in region 1 to be 5.35$\pm$0.31 keV with 
a metallicity 0.81$\pm$0.12$Z_{\sun}$, and a column density of 
0.3$\pm$0.02$\times10^{21}\mathrm{cm}^{-2}$. The temperatures 
and pressures from all regions are listed in Table~\ref{table1} 
where the region numbering follows from Fig.~\ref{fig:cimage}

\begin{table}
\begin{center}
\centering
\begin{tabular}{c c c c}
\hline
\multicolumn{1}{c}{Regions} & 
\multicolumn{1}{c}{T (keV)} &
\multicolumn{1}{c}{$\Delta$T (keV)} &
\multicolumn{1}{c}{P ($10^{-10}$ erg/$\mathrm{cm}^3$)}\\
\hline
1 (Cavity) & 5.35 & 0.31 & 3.77\\
2 & 5.58 & 0.33 & 4.40\\
3 & 3.74 & 0.12 & 4.06\\
4 & 5.75 & 0.55& 4.87\\
5 & 4.04 & 0.16& 4.51\\
6 & 3.77 & 0.18& 4.00\\
7 & 4.17 & 0.15& 5.02\\
8 & 5.73 & 0.28& 5.04\\
9 & 4.07 & 0.17& 4.50\\
10& 5.06 & 0.36& 4.56\\
\hline
\end{tabular}
\end{center}
\caption{X-ray temperatures and pressures of the ten regions 
defined in Fig.~\ref{fig:cimage}. The largest uncertainties 
in the pressure is contributed by the estimated length of
the region along the line of sight. For details see text.}
\label{table1}
\end{table}
\begin{figure}
\begin{center}
\resizebox{\hsize}{!}{\includegraphics[width=6.5cm]{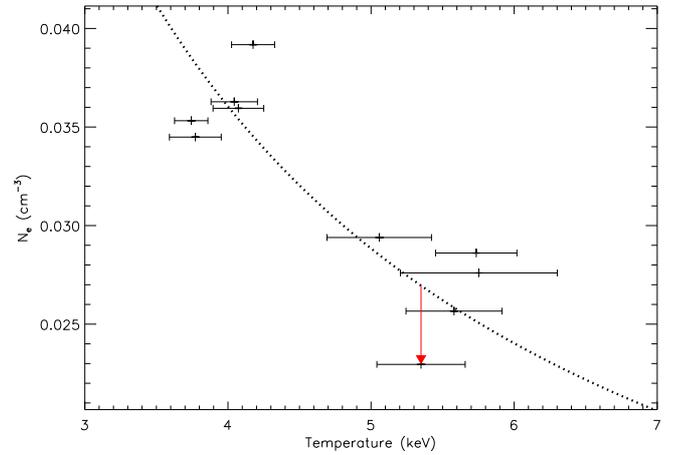}}
\end{center}
\caption{Electron density versus temperature of the ten regions 
around the X-ray cavity. We expect to have an equi-pressure in 
the central ICM, which is indicated by a dashed line. The largest 
departure from the expected pressure is exhibited by the cavity of 
our study marked by a red arrow.}
\label{fig:pressure}
\end{figure}

To calculate the pressure inside each region a geometry of the volume 
needs to be assumed, which in principle introduces the largest uncertainty 
for this work. We assume the shocked region is contained in a cylinder
whose length is defined along two hotspots with a radius, 100~kpc. This 
radius is chosen since simulations show the shock radius along 
the semi-minor axis to be roughly 100~kpc (\citealt{krause}). 
The length of each region is then the distance it extends 
along the line of sight contained in the cylinder. With this definition 
we calculated the density and the thermal pressure of the ten regions
listed in Table~\ref{table1}. We expect all ten regions to be in near 
thermal pressure equilibrium. However, our findings show that 
the region 1 contains the smallest contribution from the thermal 
pressure among ten areas. In Fig.~\ref{fig:pressure} the measured temperature 
and derived $n_e$ are plotted with the average pressure equilibrium 
in a dashed line. The statistical error of $n_e$ is very small, 
which is dominated by the geometric assumption about the volume that 
the electrons occupy. We see that much of the thermal pressure scatters 
about the expectation, however, the pressure of region 1 has the 
largest offset, roughly 16~\% lower than the expectation. 
This implies that the volume which the thermal plasma occupies has to 
be 32~\% smaller to be in pressure equilibrium with the surroundings. 

Here we continue with the hypothesis that the missing thermal pressure 
in the volume can be explained by introducing non-thermal pressure 
support. In the next section we will use this 32~\% of the volume
to calculate the relativistic plasma and magnetic field properties.

\section{Properties of the relativistic plasma in the cavity}

\begin{figure}
\begin{center}
\resizebox{\hsize}{!}{\includegraphics[width=7cm]{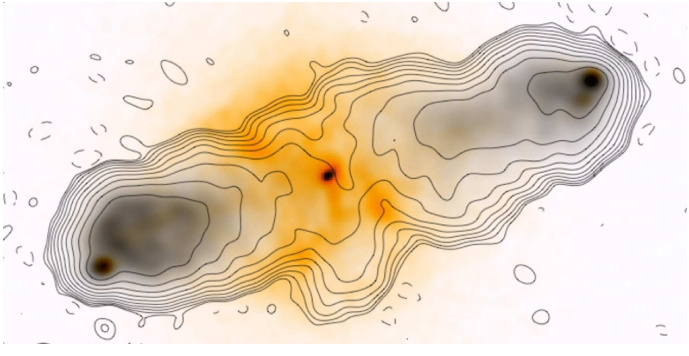}}
\end{center}
\caption{Overlay of a \emph{LOFAR} radio map as contours on 
the \emph{Chandra} image. The \emph{LOFAR} contours are taken from 
Fig.~\ref{fig:cimage} of~\citet{lofar}, and the image dimension 
is 160\arcsec$\times$80\arcsec. Notably two hotspots and the low 
surface brightness cavity of our study in the X-ray image matches 
spatially to the excess in the radio contours.}
\label{fig:radio}
\end{figure}

\begin{figure}
\begin{center}
\resizebox{\hsize}{!}{\includegraphics{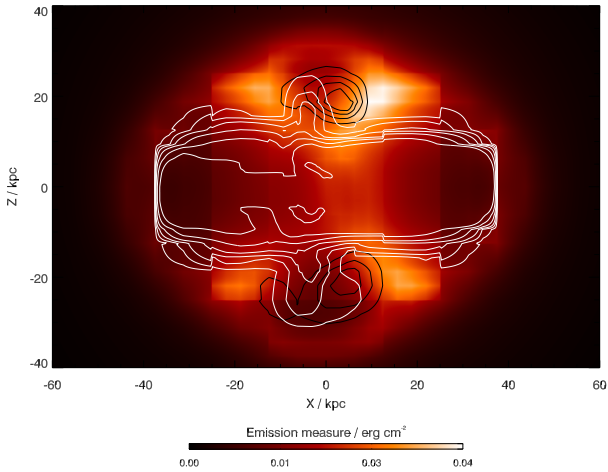}}
\end{center}
\caption{Simulated X-ray emission measure map of the interaction 
of two perpendicular jets in Cyg A. The four bright regions are 
shaped by the cavities of the first jets. 
The cavities of the main jets are also visible. Black contours show 
the volume fraction of the first jets at $2^n\times10^{-5}$. 
That of the main jets is at 1,3,6,12,24 and 48$\times10^{-3}$ in white. 
Rectangular features are due to artifacts of the nested static grid setup.
}
\label{fig:msim}
\end{figure}

We now search for an explanation of the missing cavity pressure
by a non-thermal contribution with radio observation.
We compared X-ray images of Cyg A with radio maps from 
\emph{VLA} and \emph{LOFAR} 
observations. Two \emph{VLA} observations are at 74 and 
357 MHz from~\citet{vla}, and the new commissioning \emph{LOFAR} 
data at 238 MHz is from~\citet{lofar}. The overlay with the recent 
\emph{LOFAR} observation is shown in Fig.~\ref{fig:radio}. 
It is obvious that the X-ray cavity perfectly coincides with 
the radio emission region extending out from the bridge between 
the radio lobes south of the nucleus. The highest contours
trace well the cavity region while the lowest contours encompass 
also the cavity rim. As was argued in the previous section 
the missing thermal pressure can be partly explained if 32~\% of 
the cavity volume is filled with relativistic plasma. In this 
section we calculate the relativistic plasma and magnetic field 
pressure of the cavity.

Radio observations provide information on the energy loss of 
the relativistic electrons in a magnetic field through synchrotron 
radiation. The magnetic field strength is conventionally derived 
by minimising the total energy content in the volume based on the 
equipartition argument. The magnetic field energy is proportional 
to the square of the magnetic field strength while the total electron
energy is calculated by integrating the product of its density 
and energy distribution. 
The total plasma energy of a volume can 
then be written in terms of a magnetic field strength,  
$E_{tot} \propto \alpha(1+k) LB^{-3/2} + B^2\Phi V/8\pi$ where $k$ is
the ratio of proton to electron energy, $L$ is the intensity of 
the radio emission, $\Phi$ is the filling factor describing 
the fractional volume that the magnetic field occupies, 
and $V$ is the volume of interest (e.g.~\citealt{mcnamara}). 
The corresponding pressure of the total energy in relativistic 
plasma plus magnetic field is then 13/9 $B^2/8\pi$ (\citealt{pachol}).

The radio flux density and spectral index were determined based on 
the multi-frequency data mentioned earlier.
The luminosities of the cavity are estimated to be 234 and 35 Jansky at 
74 and 327 MHz, and 55 Jansky at 238 MHz. The spectral index between 
the two extreme bands is then 1.2, and 1.24 between 74 and 238 MHz. 
Conservatively allowing an error of 50~\% on the 
74 MHz observation, we find a spectral index between 1.0 and 1.5. 
Thus we adopt 1.24 for further calculations, but we checked that 
the conclusions do not change if the spectral index is varied 
within these limits. It was noted that the spectral index of the lobe 
around the hotspots is around 0.5 (\citealt{vla}), and we expect 
the cavity spectral index to be much steeper than 0.5 
containing an older electron population.

Using the mentioned equipartition argument we calculated the plasma 
properties for given flux density, spectral index, filling factor unity,
and the lower integral limit of the synchrotron luminosity, 1 GeV.
This is reasonable given the characteristic electron energy at 74 MHz 
with a magnetic field of 17 $\mu$G is 8.9 GeV.
We find 1.18$\times10^{-11}$ erg/$\mathrm{cm}^3$ for the magnetic field 
energy density, and 1.58$\times10^{-11}$ dyne/$\mathrm{cm}^2$ for the total 
relativistic plasma pressure, and a magnetic field of 17.2 $\mu$G.
The total non-thermal pressure we calculated falls short to make 
up for the missing thermal pressure in the previous section.
However, we stress that what we calculated is the \emph{minimum} pressure. 
Uncertainties could modify our scenario, for example, a higher proton 
energy ratio such as $k=$100, would increase the non-thermal pressure 
up to 46~\% of the thermal pressure compared to 7~\% as in our case. 
Other open factors which influence results are the geometry of 
the volume element as mentioned earlier, and the lower integral limit 
for the total electron energy.

From these physical parameters and some further observables 
we can also derive information on the age of the cavity. 
One constraint is derived from the fact that the cavity filled 
with relativistic plasma is much lighter than the ambient medium, 
and we thus expect it to be buoyantly rising in the cluster 
potential (\citealt{churazov}). Assuming the buoyant velocity 
can be calculated from $\mathrm{v}_b=0.5\sqrt{r/R}\mathrm{v}_K$ 
where $\mathrm{v}_k$ is the Kepler velocity, we obtain a buoyant 
rise time of 50 Myrs for the present distance, $R$=23 kpc, 
of the bubble from the centre. 
We interpret this as an approximate upper limit of the bubble age as we 
would expect an older bubble to have moved to a larger radius.
Here the cluster was modelled with an isothermal $\beta$ 
profile~(\citealt{smith}) with core radius 29\arcsec and $\beta$=0.67, 
and the cavity volume contains negligible thermal plasma.

We can estimate the minimum age for this cavity assuming 
that electrons lose most of 
their energy through synchrotron radiation in the magnetic field. 
If we adopt the break frequency at 750 MHz determined by~\citet{specmap}
and our results for the magnetic field strength in the cavity,
we obtain a spectral age of 24 Myrs. It is interesting that the spectral
age map in Fig. 8 of~\citet{specmap} shows the X-ray cavity region as
the oldest electron population in the core of Cyg A. This fits 
together with our suggestion below that the X-ray cavity 
was created earlier than the current jets, possibly by a less 
energetic feedback event.

\section{Discussion}

We can now compare the estimated age of the X-ray cavity to the age 
of the jet and the radio lobes of Cyg A. The time duration of the 
present high power activity and the age of the radio lobes estimated 
from simulations and observations is about 30 Myrs 
(\citealt{krause,wilson2006}). This is of similar order as the 
estimated cavity age, but the limits we derived above also allow 
the cavity to be older by easily a factor of two. This provides 
some constraining information for the search of the origin of 
the cavity. One possible explanation is that the cavity is part 
of the backflow of the relativistic plasma from the hotspot region. 
It might have been broken off from the main lobes by an instability 
or as proposed in~\citet{specmap} by deflection of the backflow. 
In this case the cavity and the lobes would be expected to be of 
similar age. However, this is dynamically not easily possible 
since the growth timescale of e.g. Rayleigh-Taylor instabilities 
is $\ge$ 50~Myrs, which is much larger than the age of the 
current radio source.

However, we offer another explanation, where the radio plasma in the
cavity has been created in a previous phase of jet activity at lower 
power output. We envision a scenario where the effect of the jets is 
similar to those in the Perseus cluster, where two radio bubbles were 
inflated close to the nucleus, creating cavities in the X-ray emitting 
gas~(\citealt{hans,fabian}). If the jets have had a more 
North-South direction, and if the radio bubbles have partly been 
pushed off and partly started to rise due to buoyancy, we would 
expect to observe a cavity towards the South like the one we see. 

The enthalpy necessary to create the cavity can be used as 
an estimate of the AGN jet energy output in this past event. 
Assuming the enthalpy is given by $4PV$
(with $P$ the pressure, $V$ the volume of the cavity) we find 
an energy output of the AGN of the order of $2.2\times10^{59}$ erg.
For an activity duration of the order of 5--30 Myrs, 
the AGN power output in form of mechanical energy
amounts to $\sim$2--14$\times10^{44}$ erg/s. This is about 
two orders of magnitude below the current AGN power, well 
in line with the scenario where the jets have gently 
inflated bubbles close to the AGN rather than creating 
powerful jets that plough far out into the cluster ICM like 
at the present time.

We have further investigated this hypothesis with 3D-hydrodynamics 
simulations with the NIRVANA code (\citealt{nirvana}). 
With a static gravitational potential two jets are injected by adding 
energy, momentum and mass to a cylindrical region with a radius of 
0.2~kpc. First, a low power jet ($10^{45}$ erg/s) is injected along 
the Z-axis for 5~Myrs. After a 5~Myr break, the main jet 
(2$\times10^{47}$ erg/s) is started and evolved for 7.8~Myr. 
We show the resulting emission measure map along with contours 
of the two jet tracers through the midplane in Fig.~\ref{fig:msim}. 
Interestingly the X-ray cavities created by the first jets are 
preserved during the expansion of the main jet, and just pushed 
outwards. A characteristic feature of this model are 
the four bright regions, which appear due to the presence of 
the old cavities separating them in the middle.
This bares a striking resemblance to the X-ray observation of Cyg A. 
The jet tracers show that the plasma of the first jet was pushed 
towards the edges. Lobe plasma from the second jet and
some thermal plasma has filled the cavity and the second jet plasma
dominates the cavity pressure. The amount of the latter might in
reality be somewhat suppressed due to the stabilising effect
of magnetic fields. If in fact new plasma refills the old cavity
some of our above conclusions may be modified. 
At the simulation time of 7.8~Myr after launch of the main jets, 
they have not yet expanded to the presently observed size. 
The effects discussed here are, however, clearly demonstrated, which
includes the approximate size of the cavity. Since the simulation
employed 1.6$\times10^{59}$ erg for the first jet pair, this confirms
our order of magnitude estimate of the energy of the previous
event.

\section{Conclusions}

We have combined most of the available \emph{Chandra} X-ray data 
to construct a very detailed image of the ICM in the central region 
of Cyg A. The region around the AGN is rich in structure and one 
of the most striking features is an X-ray cavity south of the nucleus  
which is reminiscent of similar cavities in other cool cores of 
galaxy clusters. A comparison of the X-ray image with radio maps 
from \emph{VLA} and \emph{LOFAR} observations shows that the cavity 
is filled with relativistic plasma with a total enthalpy of 
$2.2\times10^{59}$ erg. The most natural explanation in our opinion 
of the radio plasma filled cavity is an origin from previous activity 
of the AGN before the present radio lobes and shocked ICM region were 
created, more than about 30 Myrs ago. The appearance of the cavity
and the X-ray morphology in the centre of Cyg A is well reproduced
by our simulation of this scenario. In this case the power of the 
jets at this earlier epoch was about 100 times lower than 
today also consistent with the simulation results. 
Thus Cygnus A would have been classified at this epoch as 
FR I radio source. If our interpretation is correct we have the 
first evidence that an FR II radio source today was an FR I source 
at an earlier time.

\begin{acknowledgements} 
MK and HB acknowledge support by the cluster of excellence 
``Origin and Structure of the Universe'' (www.universe-cluser.de). 
HB and GC acknowledge support from the DfG Transregio Program 
TR 33. GC acknowledges support from Deutsches Zentrum f\"ur 
Luft und Raumfahrt(DLR). This research has made use of 
\emph{VLA}, \emph{LOFAR}, and \emph{Chandra} data, and software 
provided by the Chandra X-ray Center (CXC), and NIRVANA code v3.5 
developed by Udo Ziegler at the Leibniz-Institut f\"ur 
Astrophysik Potsdam.
\end{acknowledgements}


\begin{thebibliography}{1} 

\bibitem[Alexander et al.(1984)]{alexander} 
Alexander, P., Brown, M.~T., Scott, P.~F., MNRAS 209, 851, 1984

\bibitem[Alexander \& Pooley(1996)]{alexander_book} 
Alexander, P., Pooley, G.~G., Cygnus A -- Study of a Radio Galaxy, 
Cambridge University Press, 1996

\bibitem[B\"ohringer et al.(1993)]{hans} 
B\"ohringer, H., Voges, W., Fabian, A.~C. et al., MNRAS, 264, 25, 1993

\bibitem[Carilli et al.(1988)]{carilli1988} 
Carilli, C.~L., Perley, R.~A., Dreher, J.~W. ApJ 334, L73, 1988

\bibitem[Carilli et al.(1991)]{specmap} 
Carilli, C.~L., Perley, R.~A., Dreher, W.~W., Leahy, J.~P., ApJ, 383, 554, 1991

\bibitem[Carilli et al.(1994)]{carilli1994} 
Carilli, C.~L., Perley, R.~A., Harris, D.~E., MNRAS 270, 173, 1994

\bibitem[Carilli et al.(1996)]{carilli1996} 
Carilli, C.~L., Barthel, P.~D., A \& AR, 7, 1, 1996

\bibitem[Churazov et al.(2000)]{churazov} 
Churazov, E., Forman, W., Jones, C., B\"ohringer, H., A \& A, 356, 788, 2000

\bibitem[Clarke et al.(1997)]{clarke} %
Clarke, D.~A., Harris, D.~E., Carilli, C.~L., MNRAS, 284, 981, 1997

\bibitem[Fabian et al.(2002)]{fabian} 
Fabian, A.~C., Sanders, J.~S., Crawford, C.~S. et al., MNRAS, 331, 369, 2002

\bibitem[Harris et al.(1994)]{harris} 
Harris, D.~E., Carilli, C.~L., Perley, R.~A., Nature, 367, 713, 1994

\bibitem[Kaiser \& Alexander(1999)]{kaiser} 
Kaiser, C.~R., Alexander, P., MNRAS, 305, 707, 1999

\bibitem[Krause(2003)]{krause03} 
Krause, M., A\&A, 398, 113, 2003

\bibitem[Krause(2005)]{krause} 
Krause, M., A\&A, 431, 45, 2005

\bibitem[Lazio et al.(2006)]{vla} 
Lazio, T.~J.~W., et al., ApJ, 642, L33, 2006

\bibitem[McKean et al.(2011)]{lofar} 
McKean, J., et al., arXiv:1106.1041, 2011

\bibitem[McNamara \& Nulsen(2007)]{mcnamara} 
McNamara, B., Nulsen, P., ARA\&A, 45, 117M, 2007

\bibitem[Pacholczyk(1970)]{pachol} 
Pacholczyk, Radio Astrophysics, Freeman, 1970

\bibitem[Smith et al.(2002)]{smith} %
Smith, D.~A., Wilson, A. S., Arnaud, K. A., et al., ApJ, 565, 195, 2002

\bibitem[Wilson et al.(2006)]{wilson2006} 
Wilson, A.~S., Smith, D.~A., Young, A.~J., ApJ, L9, 2006

\bibitem[Wilson et al.(2000)]{wilson2000} 
Wilson, A.~S., Young, A.~J., Shopbell, P.~L., ApJ, L27, 2000

\bibitem[Yaji et al.(2010)]{yaji} 
Yaji, Y., Tashiro, M., Isobe, N. et al., ApJ, 714, 37, 2010

\bibitem[Ziegler(2008)]{nirvana} 
Ziegler, U., Computer Physics Communications, 179, 227, 2008


\end{thebibliography}
\end{document}